\documentstyle[11pt]{article}
\begin{document}
\thispagestyle{empty}
\baselineskip 20pt
\rightline{CU-TP-827}
\rightline{hepth/9704098}
\
\vskip 1.5cm
\centerline{\Large\bf Witten Index and Threshold Bound States of D-Branes}
\vskip  1cm
\centerline{Piljin Yi\footnote{electronic mail: piljin@phys.columbia.edu}}
\vskip 2mm
\centerline{\it Physics Department, Columbia University, New York,  NY 10027}
\vskip 2cm

\centerline{\bf ABSTRACT}
\vskip 1cm
\begin{quote}
We consider the Witten index ${\cal I}= {\rm Tr}\, (-1)^F$ of $SU(2)$ 
Super-Yang-Mills quantum mechanics (SYMQ) with $N=16$, 8, 4 supersymmetries. 
The theory governs the interactions between a pair of D-branes under various 
circumstances, and our goal is to count the number of the threshold bound 
states directly from the low-energy effective theory. The string theory
and M theory have predicted that ${\cal I}=1$ for $N=16$, which in fact forms
an underlying hypothesis of the M(atrix) theory formulation. Also the 
consistency of conifold transitions in type II theories is known to require 
${\cal I}=0$ for $N=8$ and $4$. Here, the bulk 
contribution to ${\cal  I}$ is computed explicitly, and  for $N=16$, $8$, 
$4$,  found to be $5/4$, $1/4$, $1/4$ respectively, suggesting a common 
defect contribution of $-1/4$. We illustrate how the defect term of $-1/4$ 
may arise in the $SU(2)$ SYMQ by considering the effective dynamics along 
the asymptotic region.
\end{quote}

\newpage

In recent years, the physics of D-branes \cite{tasi} attracted a lot of 
attention. One obvious reason for this is its importance in understanding
the nonperturbative sector of the theory, but it is fair to say that 
interests in D-branes are  also due to the familiar low energy effective 
theory, namely the supersymmetric Yang-Mills theories \cite{dbound}. 
As a consequence, understanding a simple geometrical picture of D-branes 
in the context of string theory and M theory often leads to a profound and 
nontrivial results for Yang-Mills theories. For instance, the so-called
Nahm data construction of BPS monopoles \cite{nahm}, which appears quite 
mysterious at first sight, turns out to have a perfectly transparent 
explanation in terms of a network of D$p$-branes and D$(p-2)$-branes
\cite{miao}. Furthermore the same construction naturally identifies certain 
multi-monopole moduli spaces and a Coulomb phase of 3-dimensional Yang-Mills 
theories \cite{2+1}, a fact that would have been even more puzzling.

Yang-Mills theories on the world-volume is in effect a generalized moduli space
approximation to the full dynamics. That is, one identifies certain small
deformations around supersymmetric configurations and excites them in 
a nonrelativistic fashion. Because of this, it is expected to be efficient in
capturing low energy, nonrelativistic phenomena of the D-brane motion.
A typical question one can answer, for example, would be that of bound state
spectrum. There are a wealth of predictions in this regards from 
string dualities and nonperturbative effects, and the purpose of this note
is to verify such predictions in a few particular cases, all involving 
bound state at threshold.

The first plays an important role in M-theory \cite{M}. The M-theory on 
$R^{11}$ reduces to the type IIA theory upon the compactification to 
$S^1\times R^{10}$. When this happens, the Ramond-Ramond(RR)
vector field $v_\mu$ of type IIA  arises from
an off-diagonal components of the 11-dimensional metric. With respect to this
$U(1)$ gauge field, the Kaluza-Klein momentum $n$ along $S^1$ is a gauge
charge. On the type II side, there is a U-duality argument that there should
be exactly one supermultiplet that carries $v_\mu$ charge $n$ for any integer
$n$. Since the only object in the type IIA theory that carries this 
RR charge is the D-particle, such a state must be
realized as a threshold bound state of $n$ D-particles \cite{sen}. In a
recent formulation of M theory \cite{banks}, also
known as M(atrix) theory, one employs an infinite momentum 
frame where only the D-particles are the remaining degrees of freedom. A 
spectrum consistent with the low energy effective field theory were shown 
to arise if and only if there exist exactly one such bound states for each 
$n$.

The second comes from the conifold transitions \cite{conifold}
of type II theories 
compactified on a Calabi-Yau or $K3 \times T^2$. The would-be conifold 
singularity must be smoothed out by quantum effects in order for type 
II-heterotic duality to work, which in turn requires only a single
supersymmetric D-branes that wraps around a vanishing supersymmetric cycle; 
A consistent conifold transition restricts
the number of the massless particle species in the low energy field theory
at the conifold transition, and because of this, D-branes wrapping
around the supersymmetric cycles should not form a bound state at threshold. 
This conifold hypothesis has been tested in various setups. See 
Ref.~\cite{vafa} for example.

Interestingly, a single class of Yang-Mills theories appears to contain
all the relevant low energy effective theories. In the compactified cases, the 
D-branes that wrap around the supersymmetric cycles $S^3$ or $S^2$ typically 
has no zero modes along the Calabi-Yau or the $K3$ directions, so they behave 
as a D-particle in the transverse $R^4$ or the $T^2\times R^4$. 
Counting the number of supersymmetries that survive the compactification, 
it has been argued \cite{dbound,vafa}, 
one can conclude that the low energy effective theories 
are given by $N=4$ and $8$ supersymmetric Yang-Mills quantum mechanics, which
are dimensional reductions of minimal super Yang-Mills in $D=4$ and 
6.\footnote{For the $K3$ compactification, we are assuming very large $T^2$ 
(IIA) or very small $T^2$ (IIB).} The D-particle dynamics in flat $R^{10}$
is governed by $N=16$ quantum mechanics, on the other hand, which is a 
dimensional reduction  \cite{daniel,kabat} of $D=10$ Yang-Mills theory. 
In all cases, the gauge group is $U(n)=U(1)\times SU(n)$  for $n$ 
parallel D-branes, where $U(1)$ decouples and describes the free  
center-of-mass motion.

Bound states at threshold are rather difficult to probe, so usual tricks 
involve modifying the dynamics in such a way to convert the state to a 
non-threshold. One can do this by either introducing a net momentum after a
further compactification \cite{sen}, or by putting in by hand a potential 
term \cite{lowe} that 
modifies the asymptotic interaction. While this facilitates the computation 
in many cases, it is not clear exactly under what conditions such deformations
preserve the bound state structure in question. 
Seen as a purely mathematical exercise of
computing an index (and more), a qualitative modification of dynamics at 
large distances is always dangerous. Physically, one may suspect that the 
continuity of BPS spectrum is better protected when one has more 
supersymmetries, but the precise criteria are yet to be established for 
the problems at hand. Thus, we wish to address the problem of threshold
bound states more directly.

A thorough introduction to supersymmetric
Yang-Mills quantum mechanics (SYMQ) as well as some preliminary study 
was given in a recent work \cite{daniel}. Here, we will
consider the simplest case of a pair of D-branes, so the relevant gauge 
group is $SU(2)$. We obtain the extended SYMQ as a dimensional reduction of 
minimally supersymmetric Yang-Mills field theory in $D$-dimensions with 
$D=10,6,4,3$ for $N=16,8,4,2$ respectively.  Upon the dimensional reduction,
the spatial part of the gauge connection reduces to $D-1$ bosonic 
``coordinates'' $X_i^a$ in the adjoint representation. We denote the conjugate 
momenta of $X_i^a$ by $\pi^a_i$,
\begin{equation}
[\pi^a_i, X_j^b]=-i\delta^{ab}\delta_{ij}.
\end{equation}

The spinor consists of either $N$ real (for $D=10$ or $D=3$) or 
$N/2$ (for $D=6$ or $D=4$) complex adjoint fermions $\psi^a_\beta$. 
Let $\tilde\psi^a$ be the adjoint of $\psi^a$. If $\psi^a$ is real,
$\tilde\psi^a=(\psi^a)^T$ and we have,
\begin{equation}
\{\psi^a_\alpha,\psi^b_\beta\}=\delta^{ab}\delta_{\alpha\beta},
\end{equation}
while if $\psi^a$ is complex, we have
\begin{equation}
\{\psi^a_\alpha,\psi^b_\beta\}=0=\{\tilde\psi^a_\alpha,\tilde\psi^b_\beta\},
\qquad
\{\tilde\psi^a_\alpha,\psi^b_\beta\}=2\delta^{ab}\delta_{\alpha\beta}.
\end{equation}
Note that we are not using the conventional normalization for complex
fermions. In any case, it is easy to see that the canonical commutation
relations of the fermions span a $(3N)$-dimensional Clifford algebra, so
that a wavefunction in such a theory must come in as a $2^{3N/2}$-dimensional
multiplet.

For even $D$, let us define the $SO(1,D-1)$ Dirac matrices in the Weyl basis,
\begin{eqnarray}
\Gamma_0&=&1\otimes i\sigma_2 \\ 
\Gamma_i&=&\gamma_i\otimes \sigma_1 
\end{eqnarray}
with $SO(D-1)$ Dirac matrices satisfying $\{\gamma_i,\gamma_j\}=2\delta_{ij}$
and $\gamma_i^\dagger=\gamma_i$. For $D=3$, which gives  $N=2$ SYMQ, one can 
use $\Gamma_0= i\sigma_2$, $\Gamma_1=-\sigma_3$, $\Gamma_2=\sigma_1$, and
$\gamma_i=\Gamma_0\Gamma_i$. For $D=10$ and $D=3$, the $\gamma_i's$ are real 
and symmetric. With this, a  useful quantity to define  is the following 
set of bilinears,
\begin{equation}
K^a_\mu = i\epsilon_{abc}\tilde\psi^b\gamma_\mu\psi^c \qquad \mu=1,\dots,D,
\end{equation}
where we have also defined $\gamma_D\equiv -i$.
We are finally ready to write down the Hamiltonian: 
\begin{equation}
H=\frac{1}{2}\pi^a_i\pi^a_i-\frac{1}{2}X^a_iK^a_i
+\frac{1}{4}(\epsilon^{abc}X^b_i X^c_j)^2,
\end{equation}
where we suppressed the coupling constant. Because of  the local gauge 
symmetry $SU(2)$, one must impose the Gauss constraints generated by
\begin{equation}
G^a=\epsilon^{abc}X^b_i\pi^c_i-\frac{i}{2}K^a_D .
\end{equation}
Furthermore, the Hamiltonian has $N$ supersymmetries that commute with the 
gauge transformations and are generated by
\begin{equation}
Q_\alpha=\gamma_{i\alpha\beta}\psi^a_\beta\pi^a_i-\frac{1}{2}\gamma_{ij\alpha
\beta}\epsilon^{abc}\psi_\beta^aX_i^bX_j^c ,
\end{equation}
and the adjoint thereof, if $\psi$  is complex.

The Langrangian that encodes all of this is simply a dimensional reduction
\cite{daniel} of $D$-dimensional, minimally supersymmetric pure Yang-Mills,
\begin{equation}
\int dt\;{\cal L}=\int dt\;\left[\,\frac{1}{2}D_tX^a_iD_tX^a_i-\frac{1}{4}
(\epsilon^{abc}X^b_iX^c_j)^2+\frac{i}{2}\tilde\psi^aD_t\psi^a+\frac{1}{2}X^a_i
K^a_i \right],
\end{equation}
where $D_t\equiv\partial_t-iA$ with the $SU(2)$ gauge connection $A$. 
In the Hamiltonian formalism, this can be  rewritten as
\begin{equation}
\int dt\;{\cal L}=\int dt\;\left[ \pi^a_i \partial_t X^a_i+\frac{i}{2}\tilde
\psi^a\partial_t\psi^a-H+A^aG^a\right],
\end{equation}
so that the gauge connection $A^a$ plays the role of Lagrange multiplier that
imposes the Gauss constraint.

The quantity we are interested in is the number of normalizable ground
states; we are interested in counting bound states at threshold. 
However, a more convenient quantity to compute is the Witten index, 
${\cal I}$, which counts the number of bosonic ground state minus the 
number of fermionic ground states. Provided that the normalizability
condition is imposed in counting ground states, the D-brane 
predictions tell us that one must find ${\cal I}=0$ for $N=4$ and $8$, while 
${\cal I}=1$ for $N=16$. A properly regularized form of the index would be
\begin{equation}
{\cal I}=\lim_{\beta\rightarrow \infty}{\rm Tr}\,(-1)^Fe^{-\beta H},
\end{equation}
where one traces over the physical, gauge-invariant states only. The heat
kernel is easier to compute for $\beta\rightarrow 0$, so a standard trick 
is to rewrite as,
\begin{equation}
{\cal I}=\lim_{\beta\rightarrow \infty}{\rm Tr}\,(-1)^Fe^{-\beta H}
=\lim_{\beta\rightarrow 0}{\rm Tr}\,(-1)^Fe^{-\beta H}+\delta{\cal I},
\end{equation}
where the defect term is defined formally by
\begin{equation}
\delta{\cal I}
=\int_0^\infty d\beta\,\frac{d}{d\beta}{\rm Tr}\,(-1)^Fe^{-\beta H}.
\end{equation}
The bulk term ($\beta\rightarrow 0$) may contain contributions from the
continuous spectrum above the ground state, and the defect term $\delta 
{\cal I}$ corrects this. $\delta{\cal I}$ would vanish identically, either if
the spectrum were discrete or if the bosonic and the fermionic eigenvalue 
densities in the continuum sector were identical. A similar index problem 
has been studied by Sethi and Stern recently \cite{H} in the context of 
H-monopole problem, where they found vanishing $\delta{\cal I}$.
Unfortunately, this will not be the case in our problem.  

The bulk contribution can be expressed as an integral of a coincident 
heat kernel over the bosonic degrees of freedom,
\begin{equation}
\lim_{\beta\rightarrow 0}{\rm Tr}\,(-1)^Fe^{-\beta H}
=\lim_{\beta\rightarrow 0}
\int [dX]\;\langle X\vert\,{\rm tr}\, (-1)^F e^{-\beta H} {\cal P}
\,\vert X\rangle ,
\end{equation}
where we inserted the projection operator to isolate the gauge-invariant
states \cite{H},
\begin{equation}
{\cal P}=\frac{1}{8\pi^2}\oint_{S^3/Z_2} [d\eta] \,e^{i\eta^a G^a}.
\end{equation}
Note that the Haar measure we chose is that of $SO(3)$ 
instead of $SU(2)$. This is because only $SO(3)=SU(2)/Z_2$ acts faithfully;
the adjoint representation is invariant under the center $Z_2$.
The trace inside the integral is over the spinor representation of a 
$3N$-dimensional Clifford algebra spanned by the $\psi$'s.

As far as the $\beta\rightarrow 0$ behavior is concerned, we may use the heat
kernel expansion,
\begin{equation}
\langle X\vert \,e^{-\beta H}\, \vert X'\,\rangle=
\frac{1}{(2\pi\beta)^{3(D-1)/2}}
e^{-(X'-X)^2/2\beta}e^{-\beta(V+H_F)}\left(1+O(\beta)\right).
\end{equation}
The bosonic potential is denoted by $V$, while $H_F\equiv -X^a_iK^a_i/2$.
The bosonic part of the Gauss constraint rotates $\vert 
X\rangle $ by $\eta$ to $\vert X(\eta) \rangle $, and the bulk contribution 
can be rewritten as,
\begin{equation}
\lim_{\beta\rightarrow 0}\:\frac{1}{8\pi^2 (2\pi\beta)^{3(D-1)/2}}
\int [dX]\oint[d\eta]\;{\rm tr}\, (-1)^F
e^{-(X(\eta)-X)^2/2\beta}e^{-\beta\,(V+H_F)}
e^{\eta^a K_D^a/2}
\end{equation}
For generic values of $X$, $(X(\eta)-X)^2/\beta$ diverges as
$\beta\rightarrow 0$, so it is sufficient to consider small values of
$\eta$ only. Then $(X(\eta)-X)^a= \epsilon^{abc}\eta^bX^c+O(\eta^2)$.
Rescaling $\eta \rightarrow \beta \xi $, the expression is rewritten as
\begin{equation}
\lim_{\beta\rightarrow 0}\:\frac{\beta^3}{8\pi^2 (2\pi\beta)^{3(D-1)/2}}
\int [dX]\int[d\xi]\;{\rm tr}\, (-1)^F
e^{-\beta\, (\xi\times X)^2/2-\beta\,V}e^{-\beta\,(H_F-
\xi^a K_D^a/2)}.
\end{equation}
Here we suppressed terms of higher order in $\beta$, which arise from the
expansion of $(X(\eta)-X)^2$ in $\eta$ as well as from rearranging the
operators in the exponents.

Recalling that $V=-[X_i,X_j]^2/4\: (\ge 0)$ and $H_F=-X^a_iK^a_i/2$, we can 
see that the bosonic coordinates $X^a_i$ and the gauge parameters $\xi^a$ 
are on equal footing. In fact, a $SO(D)$ vector $Z_\mu^a$ with  
$\mu=1,\dots,D$, can be defined as
\begin{eqnarray}
Z_i^a&\equiv&\left(\frac{\beta}{4}\right)^{1/4}X^a_i, \qquad i=1,\dots,D-1
\nonumber\\ 
Z_D^a&\equiv&\left(\frac{\beta}{4}\right)^{1/4}\xi^a ,
\end{eqnarray}
upon which the exponents are reorganized as
\begin{equation}
-\beta\, (\xi\times X)^2/2-\beta\,V=[Z_\mu,Z_\nu]^2,
\end{equation}
and
\begin{equation}
-\beta\,H_F+\frac{\beta}{2}\,\xi^a K_D^a=\left(4\beta^3\right)^{1/4}
\frac{Z_\mu^a K_\mu^a}{2}.
\end{equation}
The measure transforms as
\begin{equation}
[dX][d\xi]=\left(\frac{\beta}{4}\right)^{-3D/4}[dZ],
\end{equation}
so that the net constant in front of the integral becomes
\begin{equation}
\frac{\beta^3}{8\pi^2 (2\pi\beta)^{3(D-1)/2}}\left(
\frac{4}{\beta}\right)^{3D/4}=\frac{1}{8\pi^2}\left(\frac{2}{\pi}
\right)^{3(D-1)/2}\left[\frac{1}{(4\beta^3)^{1/4}}\right]^{3(D-2)}.
\end{equation}
The new expression for the bulk contribution is  then
\begin{equation}
{\cal I}-\delta{\cal I}=\frac{1}{8\pi^2}\left(\frac{2}{\pi}\right)^{3(D-1)/2}
\left(\;\lim_{\lambda\rightarrow 0}\frac{1}{\lambda^{3(D-2)}}
\int[dZ]\;e^{[Z_\mu,Z_\nu]^2}\;{\rm tr}\; (-1)^Fe^{\lambda Z^a_\mu K^a_\mu/2}
\;\right).
\end{equation}
We substituted $\lambda$ for $(4\beta^3)^{1/4}$.

Since $K^a_\mu$'s are bilinear in the fermions and since there are
$3N$ fermionic coordinates, the first nontrivial trace occurs at 
the $(3N/2)$-th Taylor expansion of the fermionic part,
\begin{equation}
{\rm tr}\; (-1)^Fe^{\lambda Z^a_\mu K^a_\mu/2}=\frac{\lambda^{3N/2}}{(3N/2)!}
{\cal Q}_N(Z)+O(\lambda^{3N/2+1}) ,
\end{equation}
where ${\cal Q}_N$ is some homogeneous polynomial of order $(3N/2)$. In
computing the trace, one must be careful about the normalization of the 
fermionic coordinates. Using real fermions normalized by $\{\psi^a_\alpha, 
\psi^b_\beta\}=\delta^{ab}\delta_{\alpha\beta}$, one has 
$$
{\rm tr}\,\left(\,(-1)^F \psi^1_1\psi^2_1\cdots\psi^2_N
\psi^3_N \right) = \pm 1 .
$$
This can be understood from the fact that each $\psi^a_\alpha$ is effectively
a $SO(3N)$ Dirac matrix divided by $\sqrt{2}$. In this language $(-1)^F$ is
nothing but the chirality operator of the $3N$-dimensional Clifford algebra. 
One obtains a sensible and nontrivial form of the index formula only
if $D-2=N/2$, when the $\lambda$-dependence cancels out. But note that
this is exactly the case for $D=10,6,4,3$ where $N=16,8,4,2$ respectively.
Thus we find
\begin{equation}
{\cal I}-\delta{\cal I}=\frac{1}{8\pi^2}\left(\frac{2}{\pi}\right)^{3N/4+3/2}
\frac{1}{(3N/2)!}\int\,[dZ]\:{\cal Q}_N(Z)\,e^{[Z_\mu,Z_\nu]^2},
\end{equation}
as the form of the bulk contribution to the Witten index for all
extended $SU(2)$ SYMQ. Generalization to other gauge groups, e.g.
$SU(n)$, is straightforward.

For the simplest case  of $N=2$ ($D=3$), the computation of the polynomial is
easy enough, and  we find,
\begin{equation}
{\cal Q}_2=2\epsilon_{abc}Z^a_\mu Z^b_\nu Z^c_\lambda 
\epsilon^{\mu\nu\lambda},
\end{equation}
up to an overall sign. Note that the result is invariant
under the $SO(D=3)$ rotations on the greek indices, in addition to being 
invariant under the gauge rotations $SU(2)/Z_2$. The physical reason for this
is that we are effectively performing a Euclidean path integral with vanishing
Euclidean time $\beta$. For this reason, one can expect the $SO(D)$-invariance
to persist for other cases as well. When the fermions are complex, this is 
particularly clear. For $N=4$ ($D=4$), let us treat $\tilde\psi$ and $\psi$ 
as a completely independent pair of (2-component and complex) spinors. 
Assigning appropriate $SO(4)$ (instead of $SO(1,3)$) transformation rules
to them, the fermionic bilinears $K^a_\mu$ form a $SO(4)$ vector and  $Z_\mu^a
K_\mu^a$ is $SO(4)$-invariant; On the other hand, in taking the trace, the 
nature of these spinors are irrelevant as long as they satisfy the same set 
of canonical commutation relations. This means that ${\cal Q}_4$ has to be 
$SO(4)$ invariant, regardless of the actual symmetry properties of the complex
spinors. Indeed, we performed the computation using Maple and found,
\begin{equation}
{\cal Q}_4=480\;(\epsilon_{abc}Z^a_\mu Z^b_\nu Z^c_\lambda)^2 .
\end{equation}
A similar argument goes through for $N=8$ ($D=6$) case, although the 
invariant form of ${\cal Q}_8$ is not available at the moment.

The argument for $N=16$ ($D=10$) is a bit more involved 
because the fermions are real rather than 
complex. Suppose that we double the number of fermionic coordinates to make 
a complex Weyl spinors $\Phi^a$ and want to compute
\begin{equation}
\left[\,iZ_\mu^a\epsilon_{abc}\tilde\Phi^b\gamma_\mu\Phi^c/2\,\right]^{3N}.
\end{equation}
Treating the fermionic coordinates as classical Grassman numbers, one obtains
a homogeneous polynomial ${\cal R}(Z)$ of order $3N=48$, which multiplies
the fermionic ``volume'' form. From the same argument as
above we can see that ${\cal R}(Z)$ has to be $SO(10)$-invariant. On the 
other hand, since the $\gamma_\mu$'s are symmetric matrices for $D=10$,
we know the following decomposition occurs upon writing $\Phi=\phi+i\chi$,
\begin{equation}
Z_\mu^a\epsilon_{abc}\tilde\Phi^b\gamma_\mu\Phi^c
=Z_\mu^a\epsilon_{abc}\tilde\phi^b\gamma_\mu\phi^c+
Z_\mu^a\epsilon_{abc}\tilde\chi^b\gamma_\mu\chi^c
\end{equation}
which implies that
\begin{equation}
{\cal R}(Z)=\frac{48!}{24!\, 24!}\,\left[{\cal Q}_{16}(Z)\right]^2 .
\end{equation}
Thus, the $SO(10)$ invariance of ${\cal R}$ is inherited by ${\cal Q}_{16}$.
This concludes the proof that ${\cal Q}_N$ is $SO(D)$-invariant for all $N$.

Such rotational invariances simplify the
computation a great deal. Let us use $SO(D)$ rotations to bring $Z^a_\mu$ 
into the form,
\begin{eqnarray}
Z_1 &=& (x,a,r), \nonumber\\
Z_2 &=& (y,b,0), \nonumber\\
Z_3 &=& (z,0,0), \nonumber\\
Z_\nu &=& (0,0,0) \quad \hbox{for $\nu=4,\dots,D$}.
\end{eqnarray}
Under this ${\cal Q}_2= 3!\,2\,(zbr)$ while ${\cal Q}_4=6!\,4\, (zbr)^2$.
Because we effectively have only three nonvanishing $Z_\mu$'s in computing the
integrand, one can actually find ${\cal Q}_N$ inductively starting with
the $D=3$ or the $D=4$ result. One simply breaks up the $SO(1,D-1)$ spinor 
$\psi^a_\alpha$ to a direct sum of lower-dimensional spinors. For $D=6$, the
spinor consists of two 4-dimensional Weyl spinors, while $D=10$ needs four 
of them. The resulting decomposition into $N/4$ number of 4-dimensional
Weyl spinors (still in the adjoint representation) can be written as
\begin{equation}
\sum_{a=1}^3\sum_{j=1}^3 Z^a_j K^a_j=\sum_{a=1}^3\sum_{j=1}^3 
\left\{ Z^a_j \tilde K^a_j(1)+\cdots+Z^a_j \tilde K^a_j(N/4)\right\},
\end{equation}
where the $\tilde K^a_j(n)$'s are bilinears constructed from the $n$-th
Weyl spinors and $\gamma_j=\sigma_j$. Each piece contributes a factor
of ${\cal Q}_4$. After taking care of the combinatorics of the expansion, 
one finds
\begin{equation}
{\cal Q}_N=(3N/2)! \,2^{N/2}\,(zbr)^{N/2},
\end{equation}
which is valid for all $N\ge 2$, again up to an overall sign. We have fixed 
our convention for $(-1)^F$ so that the above choice of sign is correct.

The  evaluation of the integral is now straightforward. For $N=2$, the 
integrand is odd under $Z\rightarrow -Z$, so the integral vanishes identically.
For the other cases, we replace the $(3D)$-dimensional bosonic integral to 
those of $x,y,z,a,b,r$ multiplied by the factor of angular volumes,
\begin{equation}
\int [dZ]\quad\Rightarrow \quad
\Omega_{D-1}\Omega_{D-2}\Omega_{D-3}\int^\infty_{-\infty}
dx\,dy\,da\;\int^\infty_0 dr\, db\, dz \;r^{D-1} b^{D-2} z^{D-3},
\end{equation}
where $\Omega_{d-1}$ is the volume of a sphere $S^{d-1}$ of unit radius 
in $R^{d}$. The bosonic potential piece is also easily computed,
\begin{equation}
-[Z_\mu,Z_\nu]^2= {\cal S}\equiv
2b^2r^2+2z^2\,(a^2+b^2+r^2)+ (a^2+r^2)\,y^2+ b^2x^2-4abxy.
\end{equation}
The formula for the bulk contribution is thus reduced to a 6-dimensional
integral,
\begin{eqnarray}
{\cal I}-\delta{\cal I}&=&\frac{1}{8\pi^2}\left(\frac{2}{\pi}\right)^{3N/4+3/2}
\Omega_{N/2+1}\Omega_{N/2}\Omega_{N/2-1} \nonumber \\
& &\times 2^{N/2}\int^\infty_{-\infty}dx\int^\infty_{-\infty}dy
\int^\infty_{-\infty}da\int_0^\infty dz \int_0^\infty db \int_0^\infty dr 
\:z^{N-1}b^{N}r^{N+1}e^{-{\cal S}}.
\end{eqnarray}
We used the identity $D-2=N/2$. 

The integrations with respect to $x$, $y$,  and $z$ are Gaussian and yield a 
factor
\begin{equation}
\frac{\pi\,\Gamma(N/2)}{2^{N/2+2}\,(a^2+b^2+r^2)^{N/2}br},
\end{equation}
while the subsequent $a$-integration can be performed using a contour 
integration,
\begin{equation}
\int^\infty_{-\infty}da\,\frac{1}{(a^2+b^2+r^2)^{N/2}}=\frac{\pi\,\Gamma(N-1)
}{2^{N-2}\Gamma(N/2)^2}\,\frac{1}{\sqrt{b^2+r^2}^{\,N-1}}.
\end{equation}
Using these results, the only remaining integrations are the following one
over $b$ and $r$;
\begin{equation}
\int_0^\infty db \int_0^\infty dr\:  \frac{b^{N-1}r^N}{\sqrt{b^2+r^2}^{\,N-1}}
\,e^{-2b^2r^2}
=\frac{\pi^{1/2}\,\Gamma(N/2-1)\Gamma(N/4+1/2)}{2^{3N/4+3/2}\Gamma(N/2-1/2)},
\end{equation}
which can be found by changing variables to those of a polar coordinate 
($b=\rho\sin\theta$, $r=\rho\cos\theta$).

Putting them all together, we find
\begin{equation}
{\cal I}-\delta{\cal I}
=\frac{\Omega_{N/2+1}\Omega_{N/2}\Omega_{N/2-1}}{2^{N+3}\pi^{3N/4+1}}\,
\frac{\Gamma(N-1)\Gamma(N/2-1)\Gamma(N/4+1/2)}{\Gamma(N/2)\Gamma(N/2-1/2)}.
\end{equation}
With $\Omega_{d-1}=2\pi^{d/2}/\Gamma(d/2)$, the resulting 
bulk contributions to the Witten index are such that 
\begin{eqnarray}
{\cal I} &=& \frac{1}{4}+\delta{\cal I}\qquad  \hbox{ if } N=4, \nonumber \\
{\cal I} &=& \frac{1}{4}+\delta{\cal I}\qquad  \hbox{ if } N=8,\nonumber \\
{\cal I} &=& \frac{5}{4}+\delta{\cal I}\qquad  \hbox{ if } N=16 .
\end{eqnarray}
This is by itself very tantalizing, in that $N=16$ is the only case 
predicted to have ${\cal I}=1$ and at the same time the only case with 
the bulk contribution ${\cal I}-\delta{\cal I}$ larger than one.
Comparing this with the existing predictions for $\cal I$, one 
is lead to believe that the defect term must be $\delta{\cal I}=-1/4$ 
regardless of $N\ge 4$. In addition, we also found that ${\cal I}-\delta
{\cal I}=0$ for $N=2$.

Computation of the defect is in general a more complicated matter, 
especially because of the $L^2$ boundary condition. The defect is 
nonzero only if the spectrum is continuous. From the definition, we
have the following formal expression,
\begin{equation}
-\delta{\cal I}=\int_0^\infty d\beta \;{\rm Tr}\,(-1)^F
H e^{-\beta H}=\int_{{\cal E}>0} d{\cal E}\:\left\{\,
\rho_+({\cal E})-\rho_-({\cal E})\,\right\}, \label{defect}
\end{equation}
as an integral over all excited spectrum. We introduced $\rho_\pm({\cal E})$,
the eigenvalue densities of the Hamiltonian for bosonic and fermionic states, 
respectively. Supersymmetry ensures that
$\rho_+$ is identical $\rho_-$ on the discrete portion of the excited
spectrum, so only the continuum may contribute. The bulk contribution
computed above contains the asymmetry of the continuum (which needs not be
an integer) in addition to that of the ground states (which is an integer), 
and $\delta{\cal I}$ simply corrects this error.

One interesting aspect of SYMQ in question is its commutator potential.
The deviation from the flat direction where all $X_i$'s commute with one
another, costs more and more energy at large $X_i^a$. If one quantizes
these massive modes first (the Born-Oppenheimer approximation) treating them
as approximate harmonic oscillators at a fixed point along the flat direction,
the quantized energy grows linearly with the massless parts of $X_i^a$. 
Generic excitations of such massive modes in turn induces a linear
confining potential along the remaining would-be-massless directions, 
forcing the spectrum to be discrete \cite{daniel,kabat}. 
One obvious exception to this is when all of the massive modes are in 
their ground states, where the bosonic and the fermionic vacuum energies 
from the massive modes cancel each other. 

In short, a continuous spectrum is found in a sector where all massive modes 
are asymptotically in their ground state. To illustrate how such a sector may 
contribute a nonzero defect term, let us consider just the massless degrees 
of freedom only. After fixing the gauge $X^a_i=\delta^{a3}X_i$ and 
$\psi^a_\alpha= \delta^{a3}\psi_\alpha$, the dynamics is that of a $U(1)$ 
SYMQ, except that the Weyl group remains to act on the coordinates by $Z_2$: 
$(X_i,\psi_\alpha)\rightarrow (-X_i,-\psi_\alpha)$. This $U(1)$ theory is 
in fact an infinite coupling limit of the $SU(2)$ theory. The 
coupling $g^2$, which is suppressed above, multiplies the bosonic potential, 
which implies that only the dynamical variables at $g^2=\infty$ are along 
the classical moduli space. As the coupling grows,
a bound state, if any, would shrink to a smaller and 
smaller size set by $1/g\rightarrow 0$. Once $g^2=\infty$, thus, the
bound state must shrink to zero size and disappear from the spectrum.
The approprriate Witten index of the $U(1)$ theory has to be identically zero; 
\begin{equation}
{\cal I}_{U(1)}=0.
\end{equation}
This is obviously because we are dealing with a theory free of any interaction
whatsoever. Indeed, the Hamiltonian of this free theory is simply 
$H'=\pi_i\pi_i/2$ where $\pi_i$ is the canonical momentum of $X_i$;
a normalizable ground state has to solve the $(D-1)$-dimensional Laplace
equation and at the same time be square-integrable. But there is no
such function on the flat $R^{D-1}$.

Let $\psi_\alpha$ be the $N/2$ complex fermionic coordinates. The general 
form of the energy eigenfunction is then 
\begin{equation}
\Psi=\Psi^{(0)}(X)+\Psi^{(1)}_\alpha(X)[\psi_\alpha]+\cdots
+\Psi^{(N/2)}_{\alpha\cdots\beta}(X)[\psi_\alpha\cdots
\psi_\beta] ,
\end{equation}
where each of the $2^{N/2}$ coefficient functions is an 
eigenfunction of $H'=-\nabla^2/2$ with the Laplacian $\nabla^2$ on $R^{D-1}$.
Now the importance of the Weyl group above becomes clear. The wavefunction 
should be invariant under the $SU(2)$ gauge group, and in particular, 
$\Psi$ must be even under $(X_i,\psi_\alpha)\rightarrow 
(-X_i,-\psi_\alpha)$. Since each of the fermions is odd, the $\Psi^{(k)}$'s 
must be even for 
even $k$ and odd for odd $k$: A purely bosonic, parity-even eigenfunction
of $\nabla^2$ generates $2^{N/2-1}$ bosonic states while the parity-odd
ones generates the same number of fermionic states. 

Write the Witten index of this $U(1)$ theory again in two pieces,
\begin{equation}
{\cal I}_{U(1)}=\lim_{\beta\rightarrow 0}  \;{\rm Tr}\:(-1)^Fe^{
-\beta H'} +\delta{\cal I}_{U(1)} .
\end{equation}
In computing the bulk contribution, we use the above fact that $(-1)^F$ can
be translated to the parity operator $(-1)^P$ over ordinary functions, and 
express it in terms of a trace, ${\rm Tr}'$, over them, multiplied by the 
degeneracy factor $2^{N/2-1}$,
\begin{equation}
{\cal I}_{U(1)}-\delta{\cal I}_{U(1)}=\lim_{\beta\rightarrow 0} 2^{N/2-1}
{\rm Tr'}\;(-1)^Pe^{(\beta/2)\nabla^2}=\lim_{\beta\rightarrow 0}2^{N/2-1}\; 
\int [dX]\;\langle -X\vert \,e^{(\beta/2)\nabla^2}\,\vert X\rangle.
\end{equation}
Here, the parity operator acted on the left to convert $\langle X\vert$ to
$\langle - X\vert$.
Using ${\cal I}_{U(1)}=0$, this gives us\footnote{The integral of the heat 
kernel may appear independent of $\beta$, but this is an artifact of
taking the boundary to the infinity first. Proper thing to do is to evaluate 
the bulk and the defect contributions with the boundary at finite distance, 
say $|X|=R$, and subsequently let $R\rightarrow \infty$. For finite $R$,
the heat kernel does admit subleading terms in power of $\beta$, and the 
difference between $\beta\rightarrow 0$ and $\beta\rightarrow \infty$
generates the defect term. As far as the bulk contribution at $\beta
\rightarrow 0$ goes, however, one may take $R\rightarrow \infty$ first.}
\begin{equation}
\delta{\cal I}_{U(1)}=-\lim_{\beta\rightarrow 0}2^{N/2-1}\int [dX]\;
\frac{e^{-(X+X)^2/2\beta}}{(2\pi\beta)^{(D-1)/2}} =-\frac{2^{N/2-1}}{
2^{D-1}}=-\frac{1}{4},
\end{equation}
independent of $N$. This illustrates how the simple $Z_2$ action of the Weyl
group could induces a nontrivial defect term from the free asymptotic 
dynamics of a D-brane pair. 

Curiously enough, for $N\ge 4$, this defect term $\delta{\cal I}_{U(1)}$
appears to coincide with its $SU(2)$ counterpart $\delta{\cal I}$
expected on the basis  of the D-brane predictions. One possibility
is that this reduction to the $U(1)$ theory faithfully capture the asymmetry
of the continuous spectrum in the full $SU(2)$ problem at least for
larger dimensions. Considering the general
nature of the defect term that it is often interpretable as a boundary
contribution, and also considering that the asymptotic dynamics of the $SU(2)$
SYMQ does reduce to the $U(1)$ SYMQ divided by the Weyl group, it is certainly
plausible that one may find $\delta{\cal I}_{U(1)}=\delta{\cal I}$ under 
favorable circumstances. 

However, in this note, this computation of $\delta{\cal I}_{U(1)}$ is meant to
serve as an illustration and no more. Without more rigorous derivation of 
$\delta{\cal I}$, it is difficult to say if the reduction to the $U(1)$ 
theory is justifiable and, if so, with what boundary conditions. 
This is particularly so, in view of the peculiar case
of $N=2$ where $\delta{\cal I}_{U(1)}$ found above cannot possibly reproduce 
$\delta{\cal I}$ of the full $SU(2)$ since the bulk term vanished by itself.
A related question is whether there exist other types of continuous 
spectrum where some of the massive modes is excited. For instance, suppose
$X^1_1+iX^2_1$ is massive while $X^3_1$ is massless. If the former or its
fermionic counterpart can be excited without exciting other $X^1_i+iX^2_i$'s
in the asymptotic region, this will lead to a confining potential along 
$X^3_i$ for $i\ge 2$ but still leaves $X^3_1$ massless. Propagation along
$X^3_1$ will be free at large values of the coordinate, and one must find a 
continuous spectrum, existence of which could introduce a further defect 
contribution. We hope to return to the computation of defect contribution 
and address these subtleties.

In summary, we have obtained the bulk contribution to the Witten index 
${\cal I}$ of $SU(2)$ SYMQ. It is found to be $5/4$, $1/4$, $1/4$, $0$ 
for $N=16$, 8, 4, 2, respectively. 
The D-brane bound state spectrum is consistent with this, provided
that the defect term $\delta{\cal I}$ is equal to $-1/4$ for $N\ge 4$.
We studied the vanishing index of a reduced $U(1)$ theory to illustrate
how such a defect term may arise. Accurate computation of the defect 
contribution is still an outstanding problem.

The author is grateful to Kimyeong Lee and Brian Greene for useful 
conversations. This work is supported by U.S. Department of Energy.

\end{document}